\documentclass{INTERSPEECH2023}
\usepackage[export]{adjustbox}
\usepackage{float}
\usepackage{comment,subcaption,xcolor,algorithm,algpseudocode,multicol,multirow}
\usepackage[symbol]{footmisc}
\usepackage{graphicx,midfloat,capt-of}
\usepackage[
singlelinecheck=false 
]{caption}
\usepackage{xcolor}

\interspeechcameraready

\title{Federated Representation Learning for Automatic Speech Recognition}
\name{Guruprasad V Ramesh$^1$, Gopinath Chennupati$^2$, Milind Rao$^2$, Anit Kumar Sahu$^2$, Ariya Rastrow$^2$, Jasha Droppo$^2$}
\address{
	$^1$University of Wisconsin-Madison\\
	$^2$Amazon Alexa USA}
\email{viswanathanr@wisc.edu, chennug@amazon.com,
	milinrao@amazon.com, anitsah@amazon.com, arastrow@amazon.com, drojasha@amazon.com}

\begin{document}
	
	\maketitle
	
	\begin{abstract}
		Federated Learning (FL) is a privacy-preserving paradigm, allowing edge devices to learn collaboratively without sharing data. Edge devices like Alexa and Siri are
		prospective sources of unlabeled audio data that can be tapped to learn robust audio representations. In this work, we bring Self-supervised Learning (SSL) and FL together to learn representations for Automatic Speech Recognition respecting data privacy constraints. We use the speaker and chapter information in the unlabeled speech dataset, Libri-Light, to simulate non-
		IID speaker-siloed data distributions and pre-train an LSTM encoder with the Contrastive Predictive Coding framework with FedSGD. We show that the pre-trained ASR encoder in FL performs as well as a centrally pre-trained model and produces an improvement of 12-15\% (WER) compared to no pre-training. We further adapt the federated pre-trained models to a new language, French, and show a 20\% (WER) improvement over no pre-training.
		
	\end{abstract}

	\section{Introduction}
	Federated Learning (FL)~\cite{mcmahan2017communication, kairouz2021advances} offers collaborative training of models on decentralized data distributed across multiple devices. In FL, model updates from the participating devices are shared to a central server without compromising on user data privacy, as personal data remains intact on client devices. 
	
	A wealth of speech data is available on client devices such as smartphones and voice assistants like Alexa and Siri. This data can produce robust speech models for ASR and other downstream speech tasks. The audio data is unlabeled in nature due to the lack of
	reliable transcripts, which is more challenging in FL, to
	learn from this data in supervised fashion. Alternatively, we can learn robust representations using self-supervised learning~(SSL), which are later fine-tuned with limited transcribed data. SSL attempts in speech~\cite{oord2018representation,baevski2020wav2vec, hsu2021hubert, chen2022wavlm, mohamed2022self} show the efficacy of such a two-stage training strategy. Here, we exploit FL, which offers the privacy of audio on the devices while producing effective speech transcription models. 
	
	In this paper, we combine SSL and FL, to learn speech representations for ASR. Our strategy is two-stage: first, we use the 50K hours of unlabeled monolingual (English) speech corpus, Libri-Light \cite{kahn2020libri} to pre-train a Recurrent Neural Network Transducer (RNN-T)~\cite{graves2012sequence} encoder 
	with the FL algorithm, FedSGD \cite{mcmahan2017communication}; second, we fine-tune RNN-T (with the above pre-trained encoder) on a limited amount of transcribed audio data. 
	
	We simulate non-identical and independently distributed (non-IID) data using the speaker and chapter information of the utterances. Hereafter, this data setup is referred to as \emph{speaker-siloed} data, which is used to pretrain the LSTM encoder in a Contrastive Predictive Coding (CPC)\cite{oord2018representation} framework across a set of clients using FL. 
	We show that the federated pre-trained models are similar in performance to that of the centrally pre-trained models. We further show the efficacy of FL pre-trained models in adapting to a foreign language, French.
	
	
	\section{Related Work}
	There are a handful of attempts in literature for applying FL in speech-related tasks. Some of these applications are: ASR \cite{dimitriadis2020federated,guliani2021training,gao2022end,chennupati2022ilasr,rao2023federated}, Keyword Spotting \cite{leroy2019federated,zhang2023fedaudio}, Emotion Recognition \cite{latif2020federated,feng2022semi,zhang2023fedaudio}, and Speaker Verification \cite{granqvist2020improving}. Notably, for combining FL with SSL, the only available works include Federated self-supervised learning~(FSSL)~\cite{feng2022federated} for acoustic event detection and  \cite{gao2022federated}, where the challenges involved in combining FL \& SSL due to hardware limitations on
	the client are highlighted and a wav2vec 2.0 \cite{baevski2020wav2vec} model is trained with FL on Common-Voice Italian data \cite{ardila2019common} and fine-tuned for ASR.
	
	In this paper, we study the use of FL for producing acoustic representations using Libri-Light 50K+ hours of audio data. Our contributions are:
	\begin{itemize}
		\item To the best of our knowledge, we are the first to integrate FL for ASR at scale that produces competitive downstream ASR performance compared to regular pretraining. 
		\item Adapting monolingual federated pretrained models to a resource-constrained target language (French) resulting in an improvement of 20\% relative WER.
	\end{itemize}

\section{Methodology}
\subsection{Generating speaker-siloed data}
One of the challenges in training models in a federated setup is the diverse data distribution among the participating devices. In the context of speech assistants, the data on a client device often belongs to one speaker and have a specific linguistic and acoustic profile. Mimicking a similar setup using an open-source dataset like Libri-Light poses a challenge. Previous research attempts to integrate open-source audio data in a federated setup involved splitting the dataset based on the speaker information \cite{gao2022federated,guliani2021training,yu2021federated} but does not include a temporal notion for the data. 
Libri-Light data contains speakerID and chapterID information embedded in the unique identifier of each utterance. Here we additionally include a temporal notion by treating chapterID as a unit of time and create non-IID partitions as follows:
\begin{enumerate}
	\item Given a speech corpus, silo them based on the speakerID.
	\item Sort the siloed utterances based on the chapter information, serving as a proxy for temporal distribution.
\end{enumerate}
For example, with three clients, \textbf{C1}, \textbf{C2}, \textbf{C3}, in a round of federated pertaining, each of them contains batches that have only one speaker's utterances. \textbf{C1} contains utterances from Speaker \textbf{S1}-Chapter1, \textbf{C2} contains utterances from Speaker \textbf{S2}-Chapter1, and so on. As Libri-Light's speaker-to-utterance ratio is not uniform, the addition of this temporal notion using the chapter information ensures that different clients in a round contain data from different speakers and more closely represent a real-life scenario of federated training of speech models.

This \emph{speaker-siloed} data is used in the federated pre-training experiments. Additionally, during pre-training, we ensure that the batches generated during training are from the same speaker. Another non-trivial difference between federated and centralized pre-training is, in FL we process data once as opposed to the possibility of multiple passes in centralized pre-training. This strict constraint is to simulate the non-availability of data on the clients after a certain time threshold.

\begin{figure}[ht!]
	\centering
	\includegraphics[width=\linewidth, height=50mm]{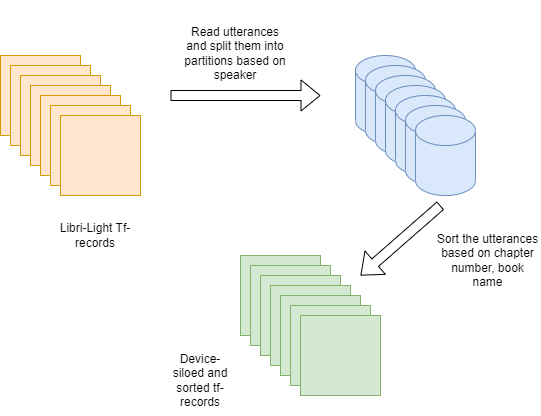}
	\captionsetup{justification=centering}
	\caption{Steps in obtaining speaker-siloed Libri-Light data for Federated pretraining}
	\label{fig:siloing}
\end{figure}

\subsection{Stage 1: Pre-training}
We compare traditional centralized pre-training with the proposed federated pretraining. \textbf{Central pre-training}: Our baseline is a model trained using the standard centralized CPC framework as outlined in Appendix \ref{cpcalgo}. Here, data is gathered and shuffled at the cloud eliminating speaker-specific distributions and running multiple epochs to learn the representations. 
\begin{algorithm}
	\caption{Federated pre-training with CPC} \label{alg:fedalgo}
	\textbf{Input}: $\mathbf{K}$ number of clients participate in each round, $\mathbf{B}$ client batch size, $\mathbf{E}$ number of local update steps and $\mathbf{C}$ number of batches per client per update-step. $\mathbf{E}$=1,$\mathbf{C}$=1 for FedSGD \\
	\textbf{Require:} Data generator $\mathbf{D_{k}}$ on $\mathbf{k}^{th}$ client.\\
	\textbf{Server:} randomly initializes weights for model, $\mathbf{w}^0$.
	\begin{algorithmic}[1]
		\For{round \(t=0,1,2,...\)}
		\State Server sends \(\mathbf{w}^t\) to the randomly selected \(\mathbf{K}\) clients.
		\For{clients \(k \in [1,\mathbf{K}]\)}
		\State $\mathbf{D_{k}}$ generates $\mathbf{C}$ batches of utterances of size $\mathbf{B}$
		\State Update the CPC model for $\mathbf{E}$ steps on the $\mathbf{C}$ batches
		\State Send the updated weights $\mathbf{w}_k^t$ to the server
		\EndFor  
		
		\State Server aggregates weights and updates its model: \hspace{2cm }$\mathbf{w}^{t+1} \gets \sum_{k=1}^{\mathbf{K}} \frac{n_k}{n}\mathbf{w^t_{k}}$
		\EndFor
	\end{algorithmic}
\end{algorithm}
\textbf{Federated pre-training}: Here, we pre-train the model with \emph{speaker-siloed} data using the FL algorithm FedSGD. Note that the hyperparameters (non-FL specific) and the network architecture are the same as central pre-training. Each of the clients involved in a round of federated pre-training updates the model based on local data and sends back weight updates to a central server. The central server accumulates and aggregates the weights and broadcasts the updated model to the participating clients of the next round (Algorithm \ref{alg:fedalgo}). Unlabelled client data is used only once in the training process as indefinite retention of data is infeasible, especially on resource-constraint devices. 

\subsection{Stage 2: Fine-tuning}
To assess the performance of the pre-trained models, both centralized and federated, we initialize the RNN-T encoder with the pre-trained model. The RNN-T encoder in our setup follows the same architecture as the CPC model. The prediction network and joint network are randomly initialized. All layers are centrally fine-tuned with transcribed audio data. 

\section{Experiments}
\subsection{Datasets}
We use Libri-Light (LL) dataset \cite{kahn2020libri} for pre-training. LL contains three parts\footnote{\url{https://github.com/facebookresearch/libri-light/tree/main/data_preparation}}: {\em small}, {\em medium}, and {\em large}, we consider the {\em large} portion, that contains $52,000$ hours of unlabeled speech data generated from $6845$ speakers. In the centralized pre-training experiments, we use all the 52K hours of the data, for FL, the same data is arranged in non-IID format, {\em speaker-siloed}, for pre-training. We further study the impact of the amount of data used in FL pre-training, for that, we use, $5K$ hours ($LL-5K$), a randomly selected subset from LL.
In fine-tuning the pre-trained models,  we use the $960$-hour train partition of the Librispeech \cite{panayotov2015librispeech}. The fine-tuned models are evaluated on the dev and test sets of Librispeech. In addition to these, we use the French data from the Multilingual Librispeech (MLS) dataset \cite{pratap2020mls} in our resource-constrained language adaptation experiments. Table \ref{table:dataset} summarizes the datasets.

\begin{table}[htp]\footnotesize
	\caption{Datasets used. For Librispeech, the clean and other splits are mentioned for dev and test} \label{table:dataset}
	\begin{tabular}{cccc}
		\hline
		\multirow{2}{*}{Dataset} & \multicolumn{3}{c}{Duration(in hours)} \\ \cline{2-4} 
		& Train & Dev & Test \\ 
		& & (clean, other) & (clean, other) \\
		\hline
		Libri-Light Large & 51934 & NA & NA \\
		Librispeech & 960.9 & 5.4, 5.4 & 5.3, 5.1 \\
		Multilingual & \multirow{2}{*}{1076.6} & \multirow{2}{*}{10} & \multirow{2}{*}{10} \\ 
		Librispeech French & & & \\
		\hline
	\end{tabular}
	\vspace{-10pt}
\end{table}
\subsection{Training Details}
The CPC~(RNN-T encoder) models in centralized and federated pre-training both consist of a $3$-layer $512$-unit feed-forward feature encoder with ReLU activation and a $6$-layer $1024$-unit unidirectional LSTM context encoder. The RNN-T model includes a $1024$-unit $2$-layer prediction network and a single dense joint layer. The inputs for both pre-training and fine-tuning models are constructed from $256$-dimensional STFT features obtained through a $25$ms window and $10$ms frame shift, combined into a final $768$-dimensional feature by concatenating three consecutive frames. All pre-training experiments are randomly initialized. The pre-trained model weights are used to initialize the RNN-T encoder in the ASR fine-tuning stage, and the prediction and joint networks are randomly initialized. The ASR model is trained with a $2500$ sentence-piece \cite{kudo2018sentencepiece} vocabulary. 
We used $48$ V100 GPUs for training the models. During pre-training, in centralized training, we ran for $130K$ steps with a bucket batch size of $[64,32,1]$. In FL, 
we vary the number of physical GPUs for the federated experiments based on the number of clients per training round involved, roughly keeping two clients per GPU and allow a maximum batch size of 8 per client. The FedSGD models are trained for 22k rounds or one pass of the entire data. On each client, SGD with a unit learning rate is used, and Adam is applied at the server with a learning rate of $1e^{-5}$. 
Fine-tuning on Librispeech is run for $100K$ steps. 

\section{Results}
\subsection{Federated versus Central Pre-training}
We show the efficacy of pre-trained (FL/central) audio representations. The RNN-T model is fine-tuned (with the pre-trained encoder) on the Librispeech $960$ hour train data. Table~\ref{table:resultsfinetune} shows the WER of the fine-tuned models on the {\em dev} and {\em test} partitions.

We compare the fine-tuned models to an RNN-T model trained without any pre-training, \emph{from-scratch}, where the encoder is randomly initialized. We also study the impact of the amount of data on pre-training with LL-5k and LL-52k datasets.
We find that all the pre-trained models (FL/central) performed better than the {\em from-scratch} model. On average, the pre-trained models show a relative WER (WERR) improvement of $11.3$\% and $14.22\%$  on {\em dev} and {\em test} sets, respectively. We observe pre-training on the larger dataset (LL-52k, see Table \ref{table:resultsfinetune}) is better in both the cases of FL and central pre-training, similar to \cite{baevski2020wav2vec,hsu2021hubert,chen2022wavlm} where centrally pre-trained models show better ASR performance with more data. We can reassure the above observation for FL pre-training to produce audio representations. 

The key observations for FL vs. central pre-training include: i) the performance of the FL pre-trained models is similar to that of the central; ii) in fact, when pre-trained on a small amount of data (LL-5k), the FL pre-trained models perform better than central, average WERR on {\em dev} and {\em test} sets is $4.76\%$ and $5.56\%$; iii) the observed superior/equal performance of the FL models is despite the single pass through the data in FL settings as opposed to multiple passes through the same data in central pre-training. We attribute the performance of the FL models to the hybrid approach of combining FL and SSL, which produces robust privacy-preserving speech representations that are useful for downstream ASR tasks.
Finally, we experimented with $48$ and $70$ clients in FL during pre-training. There is an insignificant difference in the performance of the models between the two settings. However, the impact of number of clients on the performance of the speech models can be explored in future.

\begin{table}[]
	\caption{Central pre-trained and Federated pre-trained models with RNN-T fine-tuning on train-$960$ hour Librispeech. WER results are on the dev and test sets of Librispeech. No pre-training model (trained\emph{from-scratch} on train-960 hour) is the baseline. C refers to central pre-training, and FL refer to the Federated pre-trained models. The number next to FL indicates the number of clients used in each training round.}
	\begin{tabular}{ccc}
		\hline
		\multirow{2}{*}{\begin{tabular}[c]{@{}c@{}}Pre-training\\ Setting\end{tabular}} & \multicolumn{2}{c}{Fine-tuning Word Error Rate (WER)} \\ \cline{2-3} 
		& Dev-Clean/Other & Test-Clean/Other \\ \hline
		No pretraining & 6.84/17.63 & 7.24/18.42 \\ \hline
		C, LL-5k & 6.81/17.72 & 7.3/18.2 \\
		FL48,LL-5k & 6.59/\textbf{17.02} & \textbf{6.80}/\textbf{17.42} \\
		FL70,LL-5k & \textbf{6.43}/17.17 & 6.83/17.88 \\ \hline
		C, LL-52k & 5.79/\textbf{16.29} & 6.07/\textbf{16.21} \\
		FL48,LL-52k & 5.83/16.9 & 6.16/16.34 \\
		FL70,LL-52k & \textbf{5.77}/16.4 & \textbf{6.05}/\textbf{16.21} \\ \hline
	\end{tabular}
	
	\label{table:resultsfinetune}
\end{table}

\subsection{Adaptation of SSL representations}
Self-supervised representations are beneficial for tasks with a shortage of labeled data. 
The adaptation of both multilingual and monolingual pre-trained models~\cite{conneau2020unsupervised,babu2021xls,khurana2022magic} demonstrated remarkable adaptability to other languages in speech. We explore the effectiveness of the speech representations learned from our approach (FL+SSL) in adapting to French speech.
We use French data from Multilingual Librispeech \cite{pratap2020mls}, consists of $1070$ hours of train data. The data is randomly split into two sets: $215$ hours ({\em train-215}) and $855$ hours ({\em train-855}).
We conduct two sets of experiments: Direct fine-tuning and continued pre-training followed by fine-tuning. In direct fine-tuning, the pre-trained models from Libri-Light are directly fine-tuned using the {\em train-215} data. In the other setting, we continue pre-training the LL pre-trained models using the {\em train-815} unlabeled set and only then fine-tune them using the {\em train-215} data. 
Table~\ref{tab:langadap} shows the results on the MLS French dev and test sets.

\begin{table}[htp]
	\centering\footnotesize
	\caption{Results of the language adaptation experiments on train-215 hour MLS French data. We compare no pre-training, direct fine-tuning, and continued pre-raining followed by fine-tuning. Boldface stands for the best pre-trained models.} \label{tab:langadap}
	\begin{tabular}{p{2cm}ccc}
		\hline
		Experiment & \begin{tabular}[c]{@{}c@{}}Source Pre-train\\  Setting\end{tabular} & \begin{tabular}[c]{@{}c@{}}Dev WER\\ (WERR)\end{tabular} & \begin{tabular}[c]{@{}c@{}}Test WER\\ (WERR)\end{tabular} \\ \hline
		No pre-training & - & 46.31 & 42.73 \\ \hline
		\multirow{3}{1.5cm}{Direct fine-tuning} & C, LL-52k & 37.8(18.37) & 34.6(19.02) \\
		& FL48,LL-52k & 37.82(18.33) & 34.25(19.84) \\
		& FL70,LL-5k & 43.27(6.56) & 39.69(7.11) \\ \hline
		\multirow{3}{1.5cm}{Continued pre-training + fine-tuning} & C, LL-52k & \textbf{35.92(22.43)} & \textbf{32.6(23.70)} \\
		& FL48,LL-52k & \textbf{37.63(18.74)} & \textbf{34.12(20.14)} \\
		& FL70,LL-5k & 41.48(10.42) & 37.1(13.17) \\ \hline
	\end{tabular}
\end{table}

Overall the pre-trained models perform better than from-scratch. The continued pre-training results are better than the direct fine-tuning experiments as models first adapt to the language and then to the ASR task. Another observation is that the amount of pre-training data plays a crucial role in the target language ASR performance in both the fine-tuning experiments (direct and continuous), see the significant difference in the performance of the LL-5k and LL-52k pre-trained models. FL pre-trained models are competitive with that of the centrally trained models in adapting to another language. 


\section{Conclusion}
We empirically demonstrated that FL models pre-trained in SSL style perform similarly to the centralized pre-training for the downstream ASR tasks. We employed the Contrastive Predictive Coding (CPC) framework with FedSGD at scale on a large unlabeled monolingual speech corpus, Libri-Light. The FedSGD pre-trained models also adapt to a new language, where continued pre-training on domain-specific language improves performance. In conclusion, we suggest the inflection of traditional central pre-training of audio representations to FL based pre-training to be as effective as the central case. 

In the future, we plan to extend the work to broader speech corpora, such as multilingual audio datasets and closer to real-life federated speech corpora, explore more recent SSL frameworks based on self-attention, and characterize the impact of various FL settings, such as the number of clients participating in a training round and the impact of various FL algorithms.

\section{Acknowledgements}
We acknowledge Aparna Khare and Minhua Wu in the initial discussions and the help to onboard to CPC and Anirudh Raju for his keen interest in the idea and insightful suggestions. 


\clearpage 
\appendix 

\section{Background} 
\subsection{Contrastive Predictive Coding (CPC)} \label{cpcalgo}
\begin{figure}[ht!]
	\centering
	\includegraphics[width=\linewidth, height=50mm]{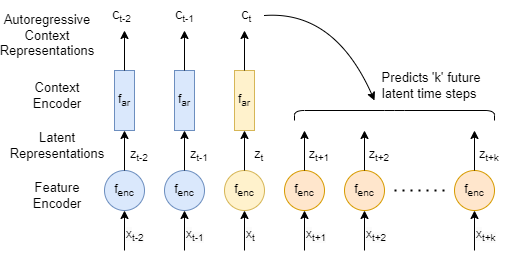}
	\captionsetup{justification=centering}
	\caption{Pretraining Architecture with CPC algorithm}
	\label{fig:cpc}
\end{figure}

CPC \cite{oord2018representation} architecture (see Figure \ref{fig:cpc}) consists of two components, the feature encoder, and the context encoder. The feature encoder \(f_{enc}\) maps the speech sequence vectors \begin{math}X=[x_1,x_2,...,x_T]\end{math} into latent representations \begin{math}Z=[z_1,z_2,...,z_T]\end{math}.  The context encoder \(f_{ar}\) is autoregressive, creating a contextualized representation \(c_{t}\)  for each time-step $t$ based on the current and previous latent representations, \(z_{t},z_{t-1},z_{t-2},...\). The pretext task of CPC is to predict a set of \(K\) future latent steps for every $ c_{t}$ $\forall t \in [1,T-K] $ with an objective to maximize the mutual information between the context \(c_{t}\) and the future latent steps $z_{t+k} 
\forall k \in [1,K] $. The entire model is optimized using a contrastive loss function \(L_C\) based on Noise Contrastive Estimation \cite{gutmann2010noise}. This loss, referred to as InfoNCE, is expressed as follows:
\begin{equation*}
	\label{eq:infonce}
	\begin{aligned}
		L_{C} &=\sum_{k=1}^{K} \frac{-1}{(T-k)}
		\sum_{t=1}^{T-k} \log \frac{e^{z_{t+k}^T(W_kc_t+b_k)/\kappa}}{\sum_{\hat{z} \in \mathbf{Z}} e^{\hat{z}^T(W_kc_t+b_k)/\kappa}},
	\end{aligned}
\end{equation*}
where $\mathbf{Z}$ is a collection of the true latent \(z_{t+k}\) and N-1 negatives sampled from other time steps in the same audio, and \(\kappa\) is the temperature term. The future latent prediction layer consists of weights $W_k$ and biases $b_k$ $ \forall k \in [1,K]$, used to predict the future latent steps. This latent prediction layer is only used in the pretraining stage. The ease of implementation and light compute footprint, compared to more recent pretraining frameworks, make CPC a great candidate for integrating it into an FL setup.  
\begin{figure}[H]
	\centering
	\includegraphics[width=40mm, height=45mm]{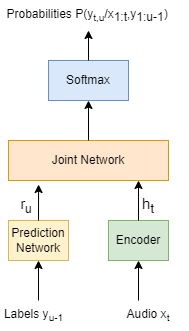}
	\captionsetup{justification=centering}
	\caption{Standard RNN-T architecture}
	\label{fig:rnnt}
\end{figure}
\subsection{Recurrent Neural Network - Transducer (RNN-T)} 
A standard RNN-T \cite{graves2012sequence} architecture  (refer to figure \ref{fig:rnnt}) consists of an encoder, a prediction network, and a joint network. The encoder processes the input speech signal \begin{math}X=[x_1,x_2,...,x_T]\end{math} and generates a frame-wise hidden representation \(h_t\) . The prediction network takes the previous non-blank tokens \(y_{u-1}\) in the output sequence predicted thus far and maps it to a hidden representation \(r_u\). The joint network combines the hidden representations of the encoder and prediction network into a softmax normalized conditional distribution over the target vocabulary augmented with the blank token.

\end{document}